\documentclass[twocolumn]{aastex631}
\usepackage{amsmath}

\shorttitle{}
\shortauthors{Zhou et al.}

\begin{document}

\title{On the Variability Features of Active Galactic Nuclei in Little Red Dots}

\correspondingauthor{Mouyuan Sun}
\email{msun88@xmu.edu.cn}

\author[0009-0005-2801-6594]{Shuying Zhou}
\affiliation{Department of Astronomy, Xiamen University, Xiamen, Fujian 361005, People’s Republic of China; msun88@xmu.edu.cn}

\author[0000-0002-0771-2153]{Mouyuan Sun}
\affiliation{Department of Astronomy, Xiamen University, Xiamen, Fujian 361005, People’s Republic of China; msun88@xmu.edu.cn}

\author[0000-0002-2420-5022]{Zijian Zhang}
\affiliation{Kavli Institute for Astronomy and Astrophysics, Peking University, Beijing 100871, People’s Republic of China}
\affiliation{Department of Astronomy, School of Physics, Peking University, Beijing 100871, People’s Republic of China}

\author[0000-0002-4765-1500]{Jie Chen}
\affiliation{Kavli Institute for Astronomy and Astrophysics, Peking University, Beijing 100871, People’s Republic of China}
\affiliation{Department of Astronomy, School of Physics, Peking University, Beijing 100871, People’s Republic of China}

\author[0000-0001-6947-5846]{Luis C. Ho}
\affiliation{Kavli Institute for Astronomy and Astrophysics, Peking University, Beijing 100871, People’s Republic of China}
\affiliation{Department of Astronomy, School of Physics, Peking University, Beijing 100871, People’s Republic of China}

\begin{abstract}
The high-redshift ($z>4$) compact sources with ``V-shaped" spectral energy distributions (SEDs), known as Little Red Dots (LRDs), are discovered by the James Webb Space Telescope and provide valuable clues to the physics of active galactic nuclei (AGNs) in the early universe. The nature of LRDs is controversial. Recently, several studies have investigated LRDs through variability, a characteristic feature of AGNs. These studies explore LRD variability by extrapolating empirical relationships from local quasars. Here, we adopt the Corona-heated Accretion-disk Reprocessing (CHAR) model, which is motivated by accretion physics and applicable to reproduce AGN conventional variability, to study the variability of $22$ LRDs in \citet{Tee2025}. Our results indicate that the observed variability in LRDs is dominated by measurement uncertainties. Within the CHAR model, the lack of variability in LRDs can be explained by two scenarios: either AGNs contribute $\lesssim30\%$ of the observed luminosities, or they are intrinsically luminous AGNs. We use simulations to demonstrate the observational requirements to effectively investigate LRDs via variability: first, a sample of about $200$ LRDs; second, each LRD has two observations separated by at least two years in the observed frame; third, the photometric uncertainty is $\leq 0.07$ mag. If the LRDs still lack variability under these conditions, the time-domain study would provide independent evidence that the accretion mode of LRDs differs significantly from low-redshift quasars.
\end{abstract}


\section{Introduction} \label{sec1:intro}
The properties of active galactic nuclei (AGNs) in the early Universe provide valuable insights into their role in reionization \citep[e.g.,][]{Dayal2020, Jiang2022, Simmonds2024}, the formation of massive black hole (MBH) seeds \citep[e.g.,][]{Fan2023, Jeon2025}, and the co-evolution of MBHs and their host galaxies \citep[e.g.,][]{Maiolino2024, Pacucci2024, Li2025}. The James Webb Space Telescope (JWST) has discovered a significant number of high-redshift ($z\gtrsim4$) compact sources characterized by ``V-shaped" spectral energy distributions (SEDs). In other words, these objects, known as Little Red Dots (LRDs), exhibit blue rest-frame UV colors and red rest-frame optical to near-infrared colors \citep[e.g.,][]{Harikane2023, Kocevski2023, Greene2024, Killi2024, Maiolino2024, Matthee2024, Wang2024, Kocevski2025, Labbe2025}. Some LRDs show broad emission lines that resemble AGN manifestations \citep[e.g.,][]{Kocevski2023, Greene2024, Wang2024}. If the broad emission lines are produced by AGNs, then these LRDs with broad lines should produce significant X-ray emission, which is observationally weak \citep[e.g.,][]{Ananna2024, Yue2024}, and suggests a surprisingly higher number density ($\sim 10^{-5}\ \mathrm{Mpc^{-3}\ mag^{-1}}$) for LRDs than the local universe extrapolation \citep[e.g.,][]{Harikane2023, Greene2024}. X-ray weakness does not necessarily contradict the AGN scenario. For example, X-ray weak quasars \citep[e.g.,][]{Wu2012} and a sample of intermediate-mass black hole AGNs with weak X-ray emission \citep[e.g.,][]{Dong2012} have been observed in the low-redshift universe; these sources can be explained by high-Eddington accretion. A similar explanation could be applied to account for the lack of X-ray emission in LRDs \citep[e.g.,][]{Inayoshi2024, Madau2024}. Still, it remains challenging to determine whether the LRDs are primarily powered by galaxies, AGNs, or a combination of both, as different SED models yield nearly identical fits \citep[e.g.,][]{Leung2024, P&G2024, Wang2024}. Therefore, new approaches, such as variability studies, should be adopted to constrain the nature of LRDs. 

As a typical characteristic of AGNs, variability is an effective probe to identify AGNs in LRDs. \citet{Kokubo2024} studied five LRDs using multi-epoch, multi-band JWST/NIRCam imaging data, finding no significant flux variations. Similarly, \citet{Zhang2025} presented magnitude variations for $\sim300$ LRDs based on multi-epoch observations from the JWST, with only eight showing significant variations beyond measurement noise on observed-frame timescales of several days to one year. Moreover, \citet{Tee2025} investigated the variability of $22$ LRDs using both Hubble Space Telescope (HST) and JWST observations, covering an observed-frame timescale of $6$ to $11$ years. This study revealed only minor magnitude variations among the $22$ LRDs. Very recently, \citet{Furtak2025} and \citet{Ji2025} found significant variations in the equivalent widths (EWs) of broad emission lines for a lensed LRD. However, no significant photometric variability exists beyond the measurement uncertainties. If the broad emission lines or UV luminosities observed in some LRDs are exclusively attributed to AGNs, then the observed lack of variability in high-redshift LRDs contradicts expectations based on extrapolations from empirical relationships derived from local quasars \citep[e.g.,][]{MacLeod2010}.   

The UV/optical variability amplitude of AGNs decreases with increasing bolometric luminosity \citep[e.g.,][]{MacLeod2010, Sun2018, Sun2020a, Suberlak2021}. However, the intrinsic bolometric luminosities of LRDs are hard to estimate because the observed SED can be fitted by different models \citep[e.g.,][]{Kocevski2023, P&G2024}. For instance, the ``AGN-only'' model with a highly dust-reddened quasar emitting at longer wavelengths, alongside its scattered emission at shorter wavelengths, results in relatively high estimates for AGN luminosity and Eddington ratio \citep[e.g.,][]{Kocevski2023, Inayoshi2024, Kokorev2024, Leung2024}. In contrast, a model that attributes the observed emission predominantly to a quasar at shorter wavelengths while considering a galaxy at longer wavelengths results in much lower estimates for luminosity and Eddington ratio \citep[e.g.,][]{Kocevski2023}. As the variability is closely related to luminosity, it can serve as an effective probe for distinguishing between different SED models.

\begin{figure}
    \centering
    \includegraphics[width=0.85\linewidth]{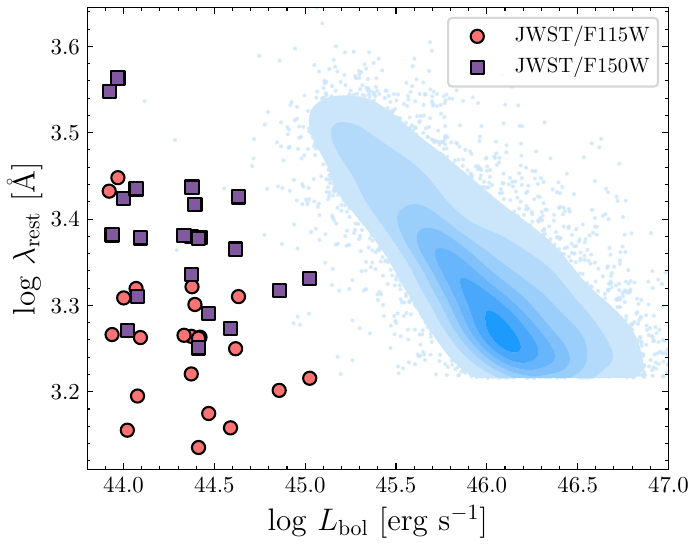}
    \caption{Rest-frame wavelengths and bolometric luminosities for LRDs and SDSS S82 quasars. The red dots and purple squares are the results for the JWST/F115W and JWST/F150W filters, respectively. The bolometric luminosities for LRDs are estimated under the assumption that the observed $L_\mathrm{1450,obs}$ are entirely due to AGN activities and without considering dust extinction for AGNs. The blue contours and dots are SDSS S82 quasars, with redshifts less than $1.9$. The rest-frame wavelengths for SDSS S82 quasars correspond to the $g$ band. The luminosities of LRDs may be one or two orders of magnitude lower than those of the SDSS S82 quasars. Hence, extrapolating the empirical variability scaling relations of SDSS S82 quasars to explain LRDs may be highly risky.}
    \label{fig1:wave-L}
\end{figure}

The lack of variability in LRDs may also be due to the dilution of their host galaxies. Based upon this assumption, \citet{Kokubo2024} and \citet{Zhang2025} found that the contamination of the host galaxy luminosity within the aperture exceeds $20\%$ of the total host galaxy luminosity in LRDs based on \citet{Burke2023} analysis of local dwarf galaxies. Moreover, \citet{Tee2025} derived the upper limit for the contribution of AGN to the total observed UV luminosity, which is about $30\%$. These estimates use the empirical relations between AGN variability amplitude and AGN properties derived from a low-redshift quasar sample of \cite{MacLeod2010} to predict the intrinsic AGN variability in LRDs. The sample analyzed by \citet{MacLeod2010} includes low-redshift but high-luminosity quasars from the Sloan Digital Sky Survey Stripe 82 (hereafter SDSS S82). It is important to note that the observed luminosities of LRDs (Figure~\ref{fig1:wave-L}) are one or two orders of magnitude lower than those of the SDSS S82 quasars. In addition, the variability parameters for high-luminosity SDSS quasars may be significantly biased due to insufficient light curve durations \citep[e.g.,][]{Kozlowski2017, Burke2021, Zhou2024, Ren2024}. Consequently, it remains uncertain whether the empirical relationship established for SDSS S82 quasars can be extrapolated to describe the high-redshift AGNs in LRDs and estimate the AGN contribution.

A physical model that can reasonably describe low-redshift quasar variability should be adopted to investigate the variability of the high-redshift AGNs in LRDs. We use an accretion-physical motivated model, the Corona-heated Accretion-disk Reprocessing \citep[CHAR;][]{Sun2020a} model, to study the variability. In the CHAR model, the corona is magnetically coupled to the standard thin disk \citep{Shakura1973}. Upon perturbation of the magnetic field in the corona, the heating rate of the accretion disk is coherently altered. As a result, the disk temperature fluctuates over time. The temperature fluctuations in response to the heating rate variations can be modeled using the thermal energy conservation law of the accretion disk. The characteristic timescale of UV/optical variability is then the thermal timescale, which is consistent with observational studies \citep[e.g.,][]{Kelly2009,Burke2021}. The framework proposed in the CHAR model can be used to reasonably describe the inhomogeneous disk of \citet{Dexter2011} if the temperature fluctuations in the inhomogeneous disk are driven by turbulence-induced heating-rate variations. In addition, the CHAR model can naturally account for other disk-driven temperature fluctuations \citep[e.g.,][]{Neustadt2022} as long as the relevant heating rate variations are adopted.

The CHAR model has been rigorously validated against the conventional variability observed in AGNs. The CHAR model predicts that the structure functions or power spectral densities of AGN UV/optical variability are nearly identical, regardless of AGN properties or rest-frame wavelengths, once the variability timescales are expressed in units of the thermal timescale \citep{Sun2020a}; this prediction agrees well with later observational studies \citep[e.g.,][]{Tang2023,Arevalo2024}. Consequently, the CHAR model successfully reproduces the observed UV/optical ensemble variability of SDSS S82 quasars, including the dependencies of variability upon timescales, wavelengths \citep[e.g., see the observational constraint from][]{Son2025}, black hole masses, luminosities \citep{Sun2020a, Sun2020b}, inter-band time delays \citep{Li2021, Chen2024a}, and the variability timescales \citep{Zhou2024, Ren2024} measured from observed-frame (up to) $20$-year long light curves \citep[e.g.,][]{Burke2021, Stone2022}. For AGNs with extremely high-cadence but long-duration \textit{Kepler} light curves, the CHAR model can reproduce their structure functions on timescales ranging from $30$ minutes to $\sim 3$ years \citep{Sun2020a}. Recently, the frequency-resolved time-lag prediction for Mrk 817 has been found to be quantitatively consistent with AGN STORM II observations \citep{Chen2024a, Chen2024b, Lewin2024}. In summary, these results demonstrate that the CHAR model provides a robust approximation for conventional quasar variability. Although not definitive, the CHAR model offers one of the best available physical frameworks for investigating the black-hole accretion driven variability in LRDs.

In this study, we use the CHAR model to study variability of $22$ LRDs in \citet{Tee2025}, whose observed-frame time intervals range from $6$ to $11$ years. The timescales probed by \citet{Tee2025} are much longer than those of other observational studies. Our analysis is based on the testable hypothesis that LRDs exhibit the same variability mechanism as normal AGNs. The aim of this manuscript is threefold. First, are the current time-domain observations of LRDs consistent with quasar conventional variability? Second, we aim to estimate the contribution of AGNs to the observed UV luminosities and intrinsic AGN luminosities under the CHAR model by comparing variability simulations with observations. Third, we propose the observational requirements for effectively studying LRDs via variability. The manuscript is organized as follows. In Section~\ref{sec2:Methodology}, we introduce the sample, model settings, and statistical method; in Section~\ref{sec3:results}, we present results and propose observational suggestions based on our simulations. Section~\ref{sec4:summary} summarizes the main conclusions. We adopt the flat $\Lambda \mathrm{CDM}$ cosmology with $H_0=70\ \mathrm{km\ s^{-1}\ Mpc^{-1}}$, $\Omega_\mathrm{m}=0.3$, and $\Omega_\Lambda=0.7$ in the manuscript.

\section{Methodology} \label{sec2:Methodology}
\subsection{LRD sample}
\citet{Tee2025} investigated the variability of an LRD sample that includes $22$ sources from EGS, GOODS-S, and Abell 2744 deep fields using HST (F125W and F160W filters) and JWST (F115W and F150W filters) observations (see their table 2).\footnote{The HST/F125W and HST/F160W filters have similar wavelength coverage to the JWST/F115W and JWST/F150W filters, so they are not distinguished in the manuscript.} The absolute magnitude ($M_\mathrm{UV}$) at the rest-frame $1450\ \mathrm{\AA}$ for the sample ranges from $-21.23$ mag to $-18.47$ mag \citep{Kocevski2025}. The observed $M_\mathrm{UV}$ for an LRD is a combination of both AGN and host galaxy contributions, and the AGN is possibly heavily obscured by dust. We calculate the observed monochromatic luminosity $L_\mathrm{1450,obs}$ according to the observed $M_\mathrm{UV}$ and luminosity distance. The bolometric luminosities for the LRDs in Figure~\ref{fig1:wave-L} are simply estimated using $L_\mathrm{1450,obs}$ and a bolometric correction factor of $3.81$ \citep{Richards2006, Shen2011}. The rest-frame wavelength for the $22$ LRDs covers the range from $1500\ \mathrm{\AA}$ to $3500\ \mathrm{\AA}$, and the rest-frame time intervals between HST and JWST observations vary from $1$ to $3$ years.

Without accounting for dust extinction, the bolometric luminosities of LRDs are at least one or two orders of magnitude lower than those of the SDSS S82 quasars used to establish the connection between AGN variability and physical properties \citep[][]{MacLeod2010}. The SDSS S82 quasars in Figure~\ref{fig1:wave-L} are taken from \citet{Sun2020b}, where only the quasars with redshifts less than $1.9$ are selected and the bolometric luminosities are calculated using the monochromatic luminosity at the rest-frame $3000\ \mathrm{\AA}$, applying a bolometric correction factor of $5.15$ \citep{Richards2006, Shen2011}. Applying the SDSS S82 quasar-based empirical variability scaling relations to LRDs may require highly risky extrapolation in luminosity. This is especially true if the physical motivation of the empirical scaling relations remains unclear and if there are significant observational biases in AGN variability measurements. Hence, we will use the accretion-physics-motivated CHAR model to simulate AGN variability in LRDs.

\subsection{Model settings} \label{subsec:Model settings}
We use the observationally validated AGN variability model, i.e., the CHAR model, to establish the AGN light curves for LRDs. Therefore, our null hypothesis is that LRDs share the same variability mechanism as normal quasars. The CHAR model has three free parameters: the dimensionless viscosity parameter $\alpha$, the black hole mass $M_\mathrm{BH}$, and the Eddington ratio $\dot{m}=L_\mathrm{bol,AGN,int}/L_\mathrm{Edd}$, where $L_\mathrm{bol,AGN,int}$ is the AGN intrinsic bolometric luminosity, and $L_\mathrm{Edd}=1.26\times 10^{38}M_\mathrm{BH}/M_\odot\ \mathrm{[erg\ s^{-1}]}$ is the Eddington luminosity. According to the CHAR model, the characteristic timescale of AGN variability scales as $\tau \propto \alpha^{-1}\lambda^{1.19}\dot{M}^{0.65}$ \citep{Sun2020a, Zhou2024}, where $\lambda$ is the rest-frame wavelength, and $\dot{M}\propto \dot{m}M_{\mathrm{BH}}$ is the absolute accretion rate; AGN light curves with similar $\tau$ share similar variability features. Note that $\dot{M}=L_\mathrm{bol,AGN,int}/(\eta c^2)$, where $\eta$ and $c$ are the radiative efficiency and the speed of light, respectively. Hence, AGN variability can be simulated once $\alpha$, $\lambda$, and $L_{\mathrm{bol,AGN,int}}$ are specified.\footnote{That is, according to the CHAR model, different reasonable choices of $M_{\mathrm{BH}}$ have negligible effects on AGN variability if $L_{\mathrm{bol,AGN,int}}$ is specified.}

The AGN intrinsic bolometric luminosity for an LRD can be estimated as follows. The observed luminosities for LRDs are a combination of contributions from AGN ($L_\mathrm{1450,AGN,obs}$) and its host galaxy. We define the AGN contribution, which indicates the contribution of the AGN to the observed luminosity, as $f_\mathrm{AGN}=L_\mathrm{1450,AGN,obs}/L_\mathrm{1450,obs}$. Then, $f_{\mathrm{AGN}}=0$ corresponds to the pure galaxy scenario for LRDs; in this case, LRDs are intrinsically non-variable. Due to dust extinction effects, $L_{\mathrm{1450,AGN,obs}}$ might be a small portion of the AGN intrinsic luminosity $L_\mathrm{1450,AGN,int}$. We introduce $\beta=L_\mathrm{1450,AGN,int}/L_\mathrm{1450,AGN,obs}$ to account for the dust extinction. For the SMC extinction curve \citep{Gordon2003}, one can find that $\mathrm{log}\ \beta\sim2A_\mathrm{V}$, where $A_\mathrm{V}$ is the magnitude of V-band extinction. In another view, the definition for $\beta$ can be inversely related to the scatter fraction parameter in ``AGN-only" SED fits of \citet{Leung2024}, which is the ratio of the scattered luminosity to the intrinsic luminosity. We set $\alpha$ to $0.4$ \citep{King2007, Sun2023}. The bolometric luminosity for AGN can be expressed as 
\begin{equation}
    L_\mathrm{bol,AGN,int}=3.81\beta f_\mathrm{AGN}L_\mathrm{1450,obs}\\,
\end{equation}
where the factor of $3.81$ is the bolometric correction factor \citep{Richards2006, Shen2011}. 

\begin{figure*}
    \centering
    \includegraphics[width=0.7\linewidth]{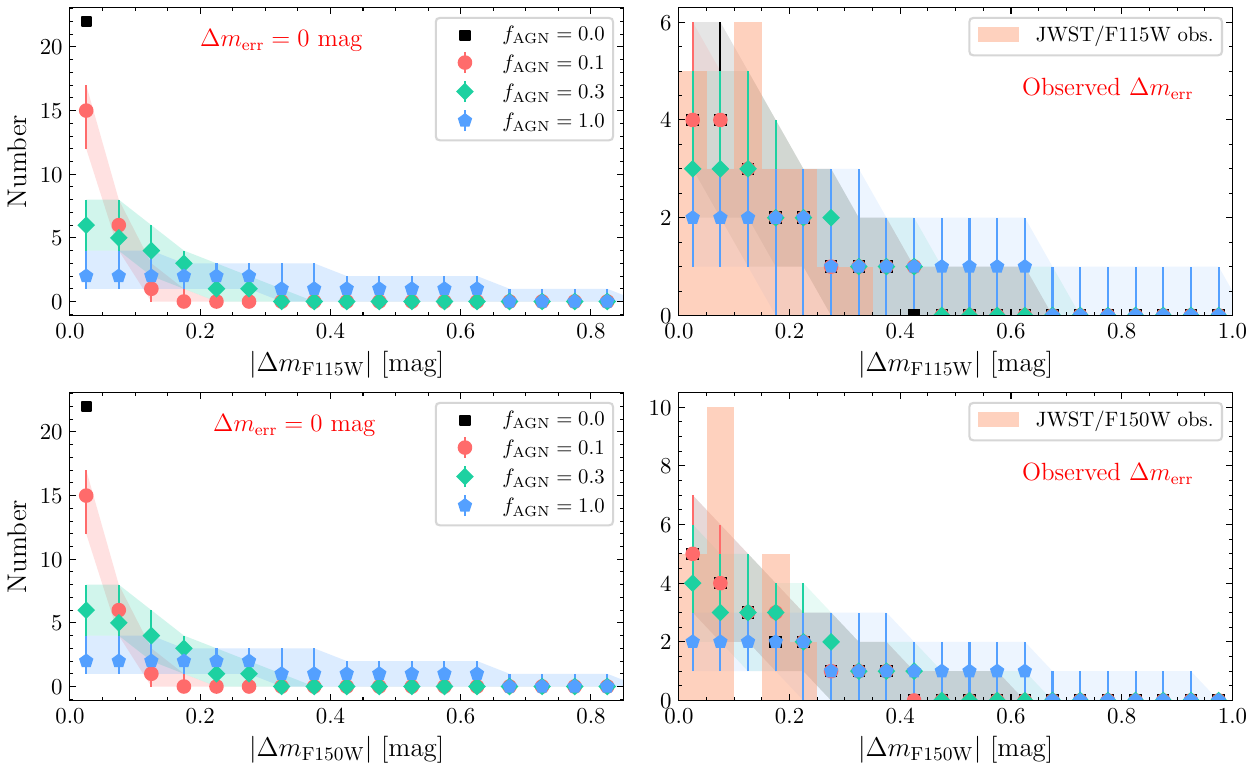}
    \caption{Distributions of the absolute magnitude variation ($|\Delta m_\mathrm{F115W}|$ or $|\Delta m_\mathrm{F150W}|$) for models with different AGN contributions ($f_\mathrm{AGN}$) and $\mathrm{log}\ \beta=0$, and real observations of \citet{Tee2025}. The first and second rows show the results for the JWST/F115W and JWST/F150W filters, respectively. In all panels, the black square, red dots, green diamonds, and blue pentagons represent the AGN contributions (i.e., $f_\mathrm{AGN}$) of $0$, $0.1$, $0.3$, and $1$, respectively. The values and error bars are the medians and $16$th--$84$th uncertainties derived from $2500$ simulations. The left panels illustrate the model results without any measurement uncertainties in $\Delta m$. In the right panels, the histograms are the observed distributions from \citet{Tee2025}, and the model results that have the same $\Delta m_\mathrm{err}$ as observations. Model results without measurement uncertainties in $\Delta m$ of different AGN contributions are distinguishable. In contrast, the observed $\Delta m$ is largely influenced by measurement uncertainties and can only provide weak constraints for $f_\mathrm{AGN}$.}
    \label{fig2:dm}
\end{figure*}

We simulate the AGN light curves ($L_\mathrm{AGN}(t)$) with different $f_\mathrm{AGN}$ and $\beta$ for each LRD using the CHAR model. The AGN contribution ($f_\mathrm{AGN}$) ranges from $0$ to $1$, with an interval of $0.1$. Motivated by the submillimeter and infrared observations \citep[e.g.,][]{Casey2025, Chen2025-dust}, the logarithmic luminosity correction factor ($\mathrm{log}\ \beta$) ranges from $0$ to $2$, with an interval of $0.2$. The corresponding magnitude of V-band extinction $A_\mathrm{V}$ ranges from $0$ to $1$ mag for the SMC extinction curve. The wavelengths for $L_\mathrm{AGN}(t)$ correspond to the rest-frame wavelengths of the JWST/F115W and JWST/F150W filters. The total light curve ($L_\mathrm{tot}(t)$) should be the combination of the variable AGN component ($L_\mathrm{AGN}(t)$) and the invariant host galaxy component ($L_\mathrm{gal}$), expressed mathematically as
\begin{equation}
    L_\mathrm{tot}(t)=L_\mathrm{AGN}(t)+L_\mathrm{gal},
\end{equation}
where $L_\mathrm{gal}=(1-f_\mathrm{AGN})\overline{L_\mathrm{tot}(t)}$, and $\overline{L_\mathrm{tot}(t)}$ is the time average value for $L_\mathrm{tot}(t)$. \citet{Tee2025} provide the observed-frame time intervals for the measured magnitude variations (see their Table 1), which allows us to determine the rest-frame time intervals for each LRD. For each LRD, we calculate the mock magnitude variations in the two bands (i.e., $\Delta m_{\mathrm{F115W}}$ and $\Delta m_{\mathrm{F150W}}$) over given rest-frame time intervals from the simulated $L_{\mathrm{tot}}(t)$, and we repeat the simulation $2500$ times to account for statistical fluctuations. To mimic the measurement uncertainties in real observations, we randomly perturb the simulated magnitude variations by adding Gaussian white noise based on the magnitude variation errors ($\Delta m_\mathrm{err}$) in Table 2 of \citet{Tee2025}.

\begin{figure*}
    \centering
    \includegraphics[width=0.75\linewidth]{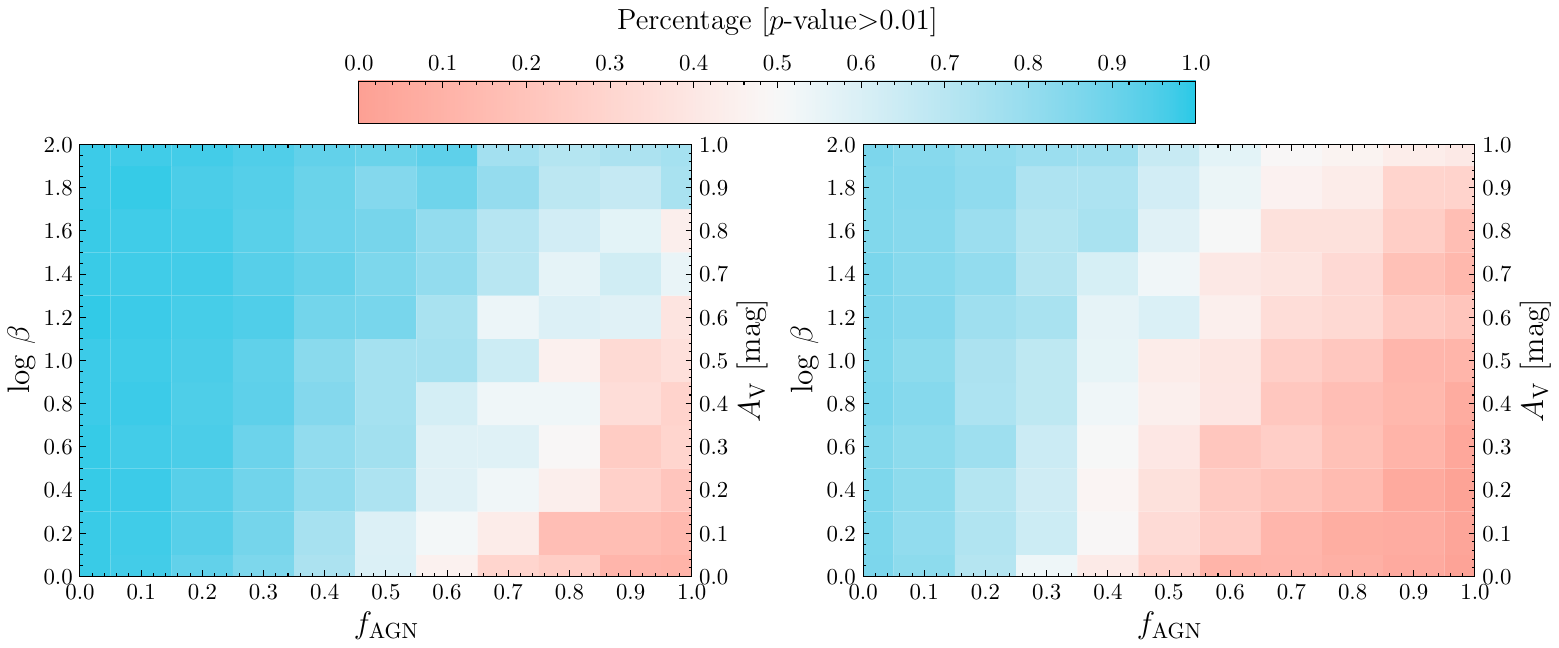}
    \caption{The percentage ($P$) of simulations with $p$-values from the A-D test between the observed and simulated magnitude variations that are greater than $0.01$. The right y-axis shows the V-band extinction ($A_\mathrm{V}$) corresponding to $\log \beta$, which is based on the SMC extinction curve. The left and right panels represent the results for the JWST/F115W and JWST/F150W filters, respectively. A lower percentage (redder in color) suggests a higher probability of rejecting the null hypothesis that the model and observations are from the same population. For $\log \beta=0$, the corresponding AGN contribution upper limit is $0.3$ for the threshold percentage $P=0.5$. The upper limits for the AGN contribution ($f_{\mathrm{AGN}}$) increase with increasing luminosity correction factor ($\beta$).}
    \label{fig3:p-value}
\end{figure*}

\subsection{Statistical approach}
For each simulation that contains mock magnitude variations for the $22$ LRDs, we construct a histogram of the absolute values of the magnitude variations ($|\Delta m|$) in either JWST/F115W or JWST/F150W filters. The histogram bins range from $0$ to $3$ mags with a bin size of $0.05$ mags. To quantitatively evaluate the differences between models with different AGN contributions ($f_{\mathrm{AGN}}$) and luminosity correction factors ($\beta$), as well as between models and observations, we conduct the Anderson-Darling test \citep[hereafter A-D test;][]{Scholz1987}. The null hypothesis for the A-D test is that the two compared samples can be generated from the same population. If the $p$-value is less than $0.01$, we reject the null hypothesis, and vice versa. We introduce a parameter $P$, representing the percentage of the $2500$ simulations with $p$-values greater than $0.01$. A lower value of $P$ indicates a higher probability that we can distinguish the two compared samples (i.e., reject the null hypothesis). The two compared samples, in our case, refer to two different distributions from simulations or observations regarding the absolute values of the magnitude variations of the $22$ LRDs. We set the threshold of $P$ to $0.5$. 

\section{Results and Discussion} \label{sec3:results}
Different AGN contributions and luminosity correction factors result in different variability of LRDs. Figure~\ref{fig2:dm} illustrates the average number of LRDs in each magnitude-variation bin based on $2500$ sets of simulations for the four $f_\mathrm{AGN}$ of $0$, $0.1$, $0.3$, and $1$ with $\mathrm{log}\ \beta=0$. 

To explore variability differences among parameters, we conduct A-D tests between the mock distribution of $f_{\mathrm{AGN}}=0.3$ and other mock distributions with different $f_{\mathrm{AGN}}$ values. For the simulated $|\Delta m|$ values without measurement uncertainties (left panels of Figure~\ref{fig2:dm}), we find that $P$ is always less than $0.5$ if $f_{\mathrm{AGN}}$ is below $0.2$ or above $0.6$. This means that the model with $f_\mathrm{AGN}=0.3$ can be distinguished from those with $f_\mathrm{AGN}<0.2$ or $f_\mathrm{AGN}>0.6$ in idealized zero measurement uncertainty cases with $\log \beta=0$. When considering the measurement uncertainties in the real observations of \citet{Tee2025}, models with different $f_\mathrm{AGN}$---including the one without AGN contribution (i.e., $f_\mathrm{AGN}=0$)---cannot be distinguished from the model with $f_\mathrm{AGN}=0.3$ (right panels of Figure~\ref{fig2:dm}). Hence, the significant measurement uncertainties in real observations are the primary limitations to constraining AGN emission in LRDs.

We now compare the mock magnitude variations with real observations. Figure~\ref{fig3:p-value} presents the corresponding $P$ values for the comparisons between models with different AGN contributions ($f_{\mathrm{AGN}}$) and luminosity correction factors ($\beta$) and real observations. Based on the values of $P$ for both the JWST/F115W and the JWST/F150W filters, our result suggests an upper limit for the AGN contribution of $0.3$ for $\log \beta=0$. This constraint is consistent with \citet{Tee2025}, albeit the model and statistical approach in our work are different from those in \citet{Tee2025}. The intrinsic bolometric luminosities for AGNs in LRDs increase with the luminosity correction factor ($\mathrm{log}\ \beta$), leading to reduced magnitude variations. Consequently, the upper limits for $f_\mathrm{AGN}$ increase with increasing $\mathrm{log}\ \beta$. That is, under the CHAR model, the observed weak LRD variability can either be explained by a powerful AGN or a weak AGN with strong galaxy contamination.  

By combining the variability and other properties of LRDs, we argue that the AGN fraction and intrinsic luminosity should not be small. First, Balmer breaks have been observed in LRDs, which can be explained by starlight by the hosts \citep[e.g.,][]{Baggen2024, Furtak2024, Wang2024} or dense clumps of neutral gas surrounding the AGN accretion disk \citep[e.g.,][]{Inayoshi2025, Ji2025}. However, if LRDs are explained by starlight alone, the inferred stellar mass would be too large to be consistent with theoretical expectations \citep[e.g.,][]{Akins2024}. Second, the EWs for the broad emission line H$\alpha$ are nearly three times larger than those for the low redshift AGNs \citep[e.g.,][]{Kokorev2024, Maiolino2025}. Note that the measured EWs are likely to be the lower limits of the intrinsic ones, and the scenario of galaxy-dominated optical emission would lead to more extreme intrinsic EWs. Third, the detailed image decomposition indicates that the galaxy contributions are negligible \citep{Chen2025}. Thus, the AGN contribution should not be very small.

\begin{figure*}
    \centering
    \includegraphics[width=0.75\linewidth]{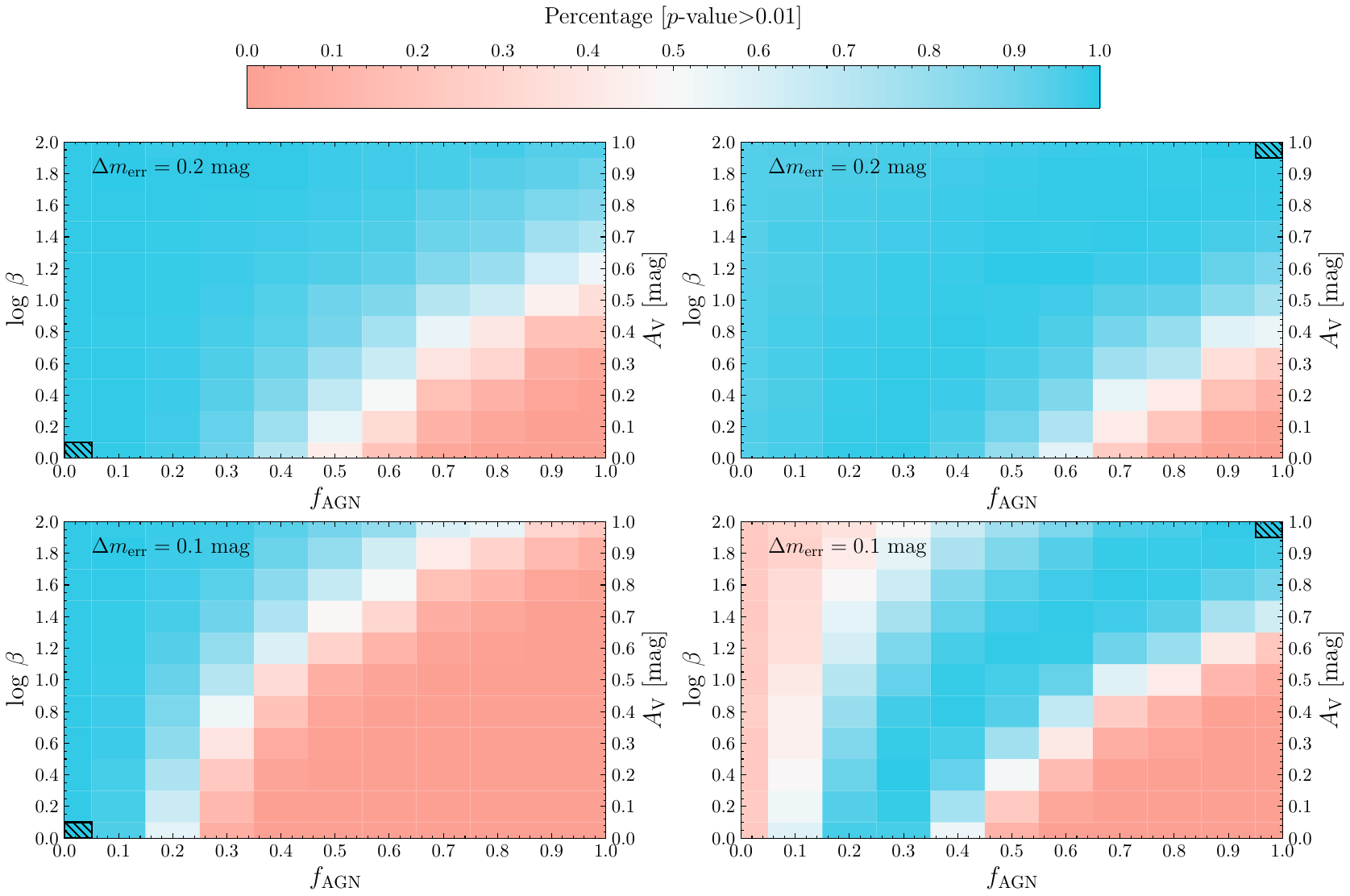}
    \caption{The values of $P$ for models with different AGN contributions and luminosity correction factors compared to a reference model. The right y-axis shows the V-band extinction ($A_\mathrm{V}$) corresponding to $\log \beta$, which is based on the SMC extinction curve. The reference models (black shaded area) are $f_\mathrm{AGN}=0$ with $\mathrm{log}\ \beta=0$, and $f_\mathrm{AGN}=1$ with $\mathrm{log}\ \beta=2$ for the left and right columns, respectively. The results are for the JWST/F150W filter, and the observed-frame time interval for each mock magnitude variation is two years. Note that the intrinsic magnitude variations are zero for $f_\mathrm{AGN}=0$ at any $\mathrm{log}\ \beta$. The magnitude variation uncertainties $\Delta m_\mathrm{err}$ are $0.2\ \mathrm{mag}$ and $0.1\ \mathrm{mag}$ for the upper and lower rows, respectively. With the current photometric accuracy ($\Delta m_\mathrm{err}\sim 0.2\ \mathrm{mag}$), we can only constrain the upper limit of the AGN contribution, even with a sample size of $200$. If the photometric accuracy is improved by a factor of two ($\Delta m_\mathrm{err}\sim 0.1\ \mathrm{mag}$), we can constrain the AGN contribution and luminosity correction factor more precisely.}
    \label{fig4:prediction}
\end{figure*}

Current variability measurements are dominated by measurement uncertainties. To obtain tighter constraints on the AGN contribution and the luminosity correction factor in the LRDs via variability, either the photometric accuracy must be greatly improved or the sample size must be significantly expanded. We first explore the potential limits to the AGN contribution and the luminosity correction factor when the LRD sample size is scaled nearly tenfold. To achieve this, we adopt the bootstrap with replacement method to generate a mock LRD sample consisting of $200$ mock LRDs from the existing $22$ LRDs. The mock LRD sample maintains the same redshift and luminosity distributions as the original $22$ LRDs. We conduct the same simulations for the mock LRD sample as in Section~\ref{subsec:Model settings}; the filter is JWST/F150W, and the time interval in the observed-frame is always two years. Figure~\ref{fig4:prediction} shows the values of $P$ for models with different $f_\mathrm{AGN}$ and $\mathrm{log}\ \beta$ compared with a reference model. The reference model is either the ``galaxy-only" case (i.e., $f_{\mathrm{AGN}}=0$; the left panels in Figure~\ref{fig4:prediction}) or the ``AGN-only" case (with $f_{\mathrm{AGN}}=1$ and $\log \beta=2$; the right panels in Figure~\ref{fig4:prediction}). With current photometric accuracy ($\Delta m_\mathrm{err}\sim 0.2\ \mathrm{mag}$), even a tenfold increase in the sample size yields weak constraints on $f_\mathrm{AGN}$ and $\mathrm{log}\ \beta$. However, if the measurement uncertainties in magnitude variations are improved to $\Delta m_\mathrm{err}\sim 0.1\ \mathrm{mag}$, as shown in the lower row of Figure~\ref{fig4:prediction}, the constraints on $f_\mathrm{AGN}$ and $\mathrm{log}\ \beta$ are rather strong. Therefore, to effectively limit both the AGN contribution and luminosity correction factor for LRDs, a sample with $\sim 200$ LRDs with the photometric uncertainty reduced to $0.1/\sqrt{2}=0.07$ mag, and $2$-year observed-frame time intervals are required\footnote{The required photometric uncertainty of the JWST/F444W filter, which corresponds to rest-frame wavelengths in the optical band, should be $\sim0.03$ mag for the 2-year observed-frame time interval. This is because AGN variability anti-correlates with wavelength.}. 

\begin{figure}
    \centering
    \includegraphics[width=0.85\linewidth]{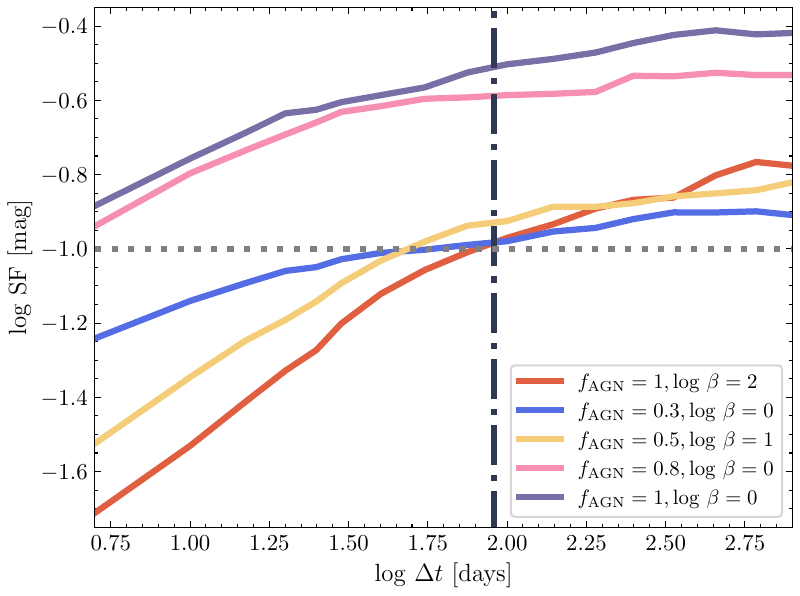}
    \caption{Examples of structure functions for an LRD at a redshift of $7.04$ with different AGN contributions and luminosity correction factors. The vertical dashed line indicates the rest-frame time corresponding to an observed time interval of two years. The horizontal dashed line corresponds to measurement uncertainties of $\Delta m_\mathrm{err}\sim 0.1\ \mathrm{mag}$. The degeneracy between $f_\mathrm{AGN}$ and $\mathrm{log}\ \beta$ leads to similar magnitude variations in some cases. In addition, the observed magnitude variations in some cases on short timescales would be dominated by measurement uncertainties.}
    \label{fig5:SF}
\end{figure}

There are certain cases, such as $f_\mathrm{AGN}=1$ with $\mathrm{log}\ \beta=2$ and $f_\mathrm{AGN}=0.3$ with $\mathrm{log}\ \beta=0$, where the magnitude variations remain indistinguishable from each other even for a $200$--LRD sample with small photometric uncertainties. This occurs for a couple of reasons. First, the degeneracy between the AGN contribution ($f_\mathrm{AGN}$) and the luminosity correction factor ($\mathrm{log}\ \beta$) results in similar mock magnitude variations in some cases. Figure~\ref{fig5:SF} illustrates the degeneracy between $f_\mathrm{AGN}$ and $\mathrm{log}\ \beta$ by presenting the structure functions (SFs) of mock light curves in the JWST/150W filter for an LRD at a redshift of $7.04$. The SF at rest-frame time interval $\Delta t$ is defined as
\begin{equation}
    \mathrm{SF}(\Delta t) = 0.74\mathrm{IQR}(\Delta m),
\end{equation}
where $\mathrm{IQR}(\Delta m)$ is the interquartile range from $25\%$ to $75\%$ of $\Delta m$. The SFs for $f_\mathrm{AGN}=1$ with $\mathrm{log}\ \beta=2$, $f_\mathrm{AGN}=0.3$ with $\mathrm{log}\ \beta=0$, and $f_\mathrm{AGN}=0.5$ with $\mathrm{log}\ \beta=1$ are similar. This is due to the fact that in the CHAR model, the variability amplitude $\sigma\propto L_\mathrm{bol, AGN, int}^{-0.25}$, which is consistent with observations \citep[e.g.,][]{MacLeod2010}. Thus, compared with the case of $\mathrm{log}\ \beta=0$, the intrinsic AGN variability amplitude for $\mathrm{log}\ \beta=2$ decreases by $100^{0.25}\approx3$, which can be balanced by a change in $f_\mathrm{AGN}$ of a factor about $3$. Second, if the intrinsic magnitude variations are smaller than measurement uncertainties, the observed variations are dominated by measurement uncertainties. Therefore, improving the photometric accuracy and increasing the observed time intervals for measuring intrinsic magnitude variations is essential. 

\citet{Zhang2025} have provided a promising LRD sample for exploring their nature via variability. The sample includes $\sim 300$ LRDs, with a median photometric uncertainty of $\sim 0.07$ mag. The observation time intervals for this sample (whose median value is $\sim 100$ days) are not long, and the observed variability is also affected by photometric uncertainties as confirmed by our simulations. Hence, the constraints of AGN luminosity from this sample are similar to \cite{Tee2025}. Upcoming observations for the sample of \cite{Zhang2025} can better probe LRD variability. 

Follow-up spectroscopic confirmations of photometrically selected LRDs indicate that $20$--$30\%$ of them are pure galaxies or brown dwarfs \citep[e.g.,][]{Greene2024, Kocevski2025}. To investigate how the presence of galaxies and brown dwarfs affects the constraints on $f_\mathrm{AGN}$ and $\mathrm{log}\ \beta$, we conducted the following test. First, we randomly select $30\%$ LRDs from our mock $\sim200$ LRD sample and set their intrinsic magnitude variation to zero for the case of $f_\mathrm{AGN}=1$ with $\mathrm{log}\ \beta=2$. Second, we add Gaussian white noise with the measurement uncertainty of $\Delta m_\mathrm{err}\sim 0.1\ \mathrm{mag}$ to the intrinsic magnitude variations. Third, we repeat the aforementioned statistical analysis for this modified case. Compared with the lower right panel of Figure~\ref{fig4:prediction}, the constraints for the AGN contribution are $10\%$--$30\%$ worse in this modified scenario. More recently, \citet{Zhang2025-NLLRD} find that $\sim 20\%$ of the photometrically selected LRDs exhibit fake ``V-shape'' SEDs due to strong line emission \citep[also see][]{Hainline2025, Hviding2025}. Thus, it would be ideal to establish a sample of LRDs with real ``V-shape'' SEDs through spectral identification and study their variability further.

We emphasize that the aforementioned observational requirements are based on the CHAR model, which describes the conventional variability of normal quasars. If LRDs continue to exhibit weaker-than-simulated variability under the proposed conditions, this can be considered as independent evidence that their time-dependent accretion processes differ significantly from those of quasars exhibiting conventional variability if the galaxy contribution is small. For instance, LRDs are often proposed to be powered by super-Eddington accretion due to their particular spectral signatures \citep[e.g., X-ray weakness; see,][]{Inayoshi2024, Madau2024, Pacucci2024-xray}; this scenario has also been speculated to be responsible for the lack of variability in LRDs \citep[e.g.,][]{Inayoshi2024, Zhang2025}. However, we caution that the super-Eddington accretion scenario does not guarantee low variability, since the viscous timescale of the puffed-up super-Eddington accretion disk \citep[e.g.,][]{Abramowicz1988}, which governs the accretion rate fluctuations, can be orders of magnitude shorter than that of the standard thin disk \citep{Shakura1973}. In summary, future observations with our proposed observational requirements, along with variability models based on black-hole accretion, can provide independent constraints on the accretion mode of LRDs.

\section{Summary} \label{sec4:summary}
We have adopted the accretion-physics-motivated CHAR model to simulate LRD variability, under the null hypothesis that their variability mechanisms align with those of normal quasars. The analysis has revealed that the observed lack of magnitude variations for LRDs in \citet{Tee2025} is dominated by measurement uncertainties and can be explained either by a powerful AGN with conventional variability or by a weak one with strong galaxy contamination. To further effectively study LRDs via variability, we recommend increasing the sample size to $\sim200$ LRDs, each with two observations separated by observed-frame time intervals of longer than two years. Additionally, the photometric uncertainty should be $\lesssim 0.07\ \mathrm{mag}$. The current JWST observations have already achieved the required sample size, and some LRDs have achieved the required photometric accuracy \citep{Kokubo2024, Zhang2025}, albeit with short time intervals. The upcoming multi-epoch JWST data, along with our recommended observational requirements, should be able to better understand the accretion mode of LRDs through variability. 

\section*{Acknowledgments}
We thank the referee for constructive comments that improved the manuscript. We acknowledge support from the National Key R\&D Program of China (2023YFA1607903 and 2022YFF0503401). S.Y.Z. and M.Y.S. acknowledge support from the National Natural Science Foundation of China (12322303) and the Fundamental Research Funds for the Central Universities (0620ZK1270). L.C.H. acknowledges support from the National Science Foundation of China (12233001).

\software{Matplotlib \citep{matplotlib}, Numpy \citep{2020NumPy-Array}, Scipy \citep{2020SciPy-NMeth}}

\bibliographystyle{aasjournal}
\bibliography{ref.bib}

\end{document}